\documentclass[%
 reprint,
 amsmath,amssymb,
 aps,
prb,
floatfix,
]{revtex4-2}

\usepackage{graphicx}
\usepackage{dcolumn}
\usepackage{bm}
\usepackage[dvipsnames]{xcolor}
\usepackage{soul}
\usepackage{comment}
\sethlcolor{yellow}

\usepackage{verbatim}
\usepackage{ulem}

\begin{document}

\newcommand{\detailtexcount}[1]{%
  \immediate\write18{texcount -merge -sum -q #1.tex output.bbl > #1.wcdetail }%
  \verbatiminput{#1.wcdetail}%
}

\newcommand{\quickwordcount}[1]{%
  \immediate\write18{texcount -1 -sum -merge -q #1.tex output.bbl > #1-words.sum }%
  \input{#1-words.sum} words%
}

\newcommand{\quickcharcount}[1]{%
  \immediate\write18{texcount -1 -sum -merge -char -q #1.tex output.bbl > #1-chars.sum }%
  \input{#1-chars.sum} characters (not including spaces)%
}

\preprint{APS/123-QED}
\title{Revealing the Microscopic Mechanism of Displacive Excitation of Coherent Phonons in a Bulk Rashba Semiconductor}

\author{P. Fischer\textsuperscript{1}}
\author{J. Bär\textsuperscript{1}}
\author{M. Cimander\textsuperscript{1}}
\author{L. Feuerer\textsuperscript{1}}
\author{V. Wiechert\textsuperscript{1}}
\author{O. Tereshchenko\textsuperscript{2}}
\author{D. Bossini\textsuperscript{1}}%
\affiliation{%
 \textsuperscript{1}Department of Physics and Center for Applied Photonics, University of Konstanz, D-78457 Konstanz, Germany\\
 \textsuperscript{2}Rzhanov Institute of Semiconductor Physics, Siberian Branch, Russian Academy of Sciences\\
}%

\date{\today}

\begin{abstract}
\noindent{}Changing the macroscopic properties of quantum materials by optically activating collective lattice excitations has recently become a major trend in solid state physics.
One of the most commonly employed light-matter interaction routes is the displacive mechanism.
However, the fundamental contribution to this process remains elusive, as the effects of free-carrier density modification and raised effective electronic temperature have not been disentangled yet.
Here we use time-resolved pump-probe spectroscopy to address this issue in the Rashba semiconductor BiTeI.
Exploring the conventional regime of electronic interband transitions for different excitation wavelengths as well as the barely accessed regime of electronic intraband transitions, we answer a long-standing open question regarding the displacive mechanism: the lattice modes are predominantly driven by the rise of the effective electronic temperature.
In the intraband regime, which allows to increase the effective carrier temperature while leaving their density unaffected, the phonon coherence time does not display significant fluence-dependent variations.
Our results thus reveal a pathway to displacive excitation of coherent phonons, free from additional scattering and dissipation mechanisms typically associated with an increase of the free-carrier density.
\end{abstract}

\maketitle


\section{\label{sec:level1}Introduction}
The generation of coherent phonons through laser pulses has recently been established as a means to both control \cite{Kim12,Nova:2016,Nov19,Fausti2011} and probe \cite{Wall2012,Mertens2023} the macroscopic phases of quantum materials.
Impulsive processes are one of the most popular approaches to the optical excitation of coherent lattice dynamics.
In particular, they allow to excite longitudinal optical (LO) phonons which are symmetrically inaccessible to resonant dipolar excitation.
In general, a distinction is drawn between two methods:
If light couples to the electronic system non-resonantly, coherent phonons are induced by impulsive stimulated Raman scattering.
On the other hand, if the optical stimulus is resonant with the electronic system, displacive excitation of coherent phonons (DECP) is realized \cite{Che90,Cho90}.
The established theory of DECP \cite{Zei92} describes the generation of coherent lattice modes in terms of two contributions, namely the increase of the charge-carrier density  $n_\textnormal{c}$ and the rise in the effective temperature of the electronic subsystem $T_\textnormal{e}$.
Crucially, so far neither theory nor experiment could disentangle these two contributions.
From an experimental point of view, typical DECP experiments involve pump photon energies resonant with electronic interband transitions.
Therefore, in the absorption process both $n_\textnormal{c}$ and  $T_\textnormal{e}$ are increased.
However, modern pulsed lasers can operate in the mid-infrared regime, with photon energies lower than the band gap of most solids.
These light sources can therefore be employed to pump electronic intraband transitions selectively, which raises only $T_\textnormal{e}$ and keeps $n_\textnormal{c}$ constant.
Hence, mid-infrared lasers offer the possibility to address the fundamental question about the microscopic mechanism of DECP, which has been unresolved for more than 30 years \cite{Zei92}.
\\\\
We study the excitation of coherent phonons in BiTeI, a prototypical Rashba semiconductor, via pump-probe time-domain reflection spectroscopy.
The band structure of this material is ideal for a quantitative comparison of interband and intraband DECP.
Due to the narrow band gap, BiTeI absorbs light well into the near-infrared spectral range via electronic interband transitions.
The high intrinsic carrier density combined with the huge Rashba spin splitting ($\Delta \varepsilon\approx$ 100~meV) of the lowest conduction band allows intraband transitions in the mid-infrared regime \cite{Mak14}.
\\\\
Our experiments reveal that in the interband regime the phonon coherence time is not affected up to an optically doped carrier density of $n_\textnormal{c}\approx1\cdot10^{21}$~cm$^{-3}$.
These findings demonstrate that BiTeI is a surprisingly robust material for coherent phononics.
This result is remarkable, considering that in other semiconductors (GaAs, GaP, InAs, InSb) the coherence time is typically suppressed by carrier densities orders of magnitude smaller than in BiTeI~\cite{Kut92,Mis01,Yee02,Has99}.
Strikingly in the intraband regime the amplitude of the phonons linearly increases with the laser fluence, even though $n_\textnormal{c}$ remains constant.
Even more tantalizing, the phonon coherence time remains unaffected, even if the laser fluence is set to a value one order of magnitude higher than in the case electronic interband pumping.
\section{\label{sec:Setup}{Materials and Methods}}
\begin{figure}[t]
\includegraphics{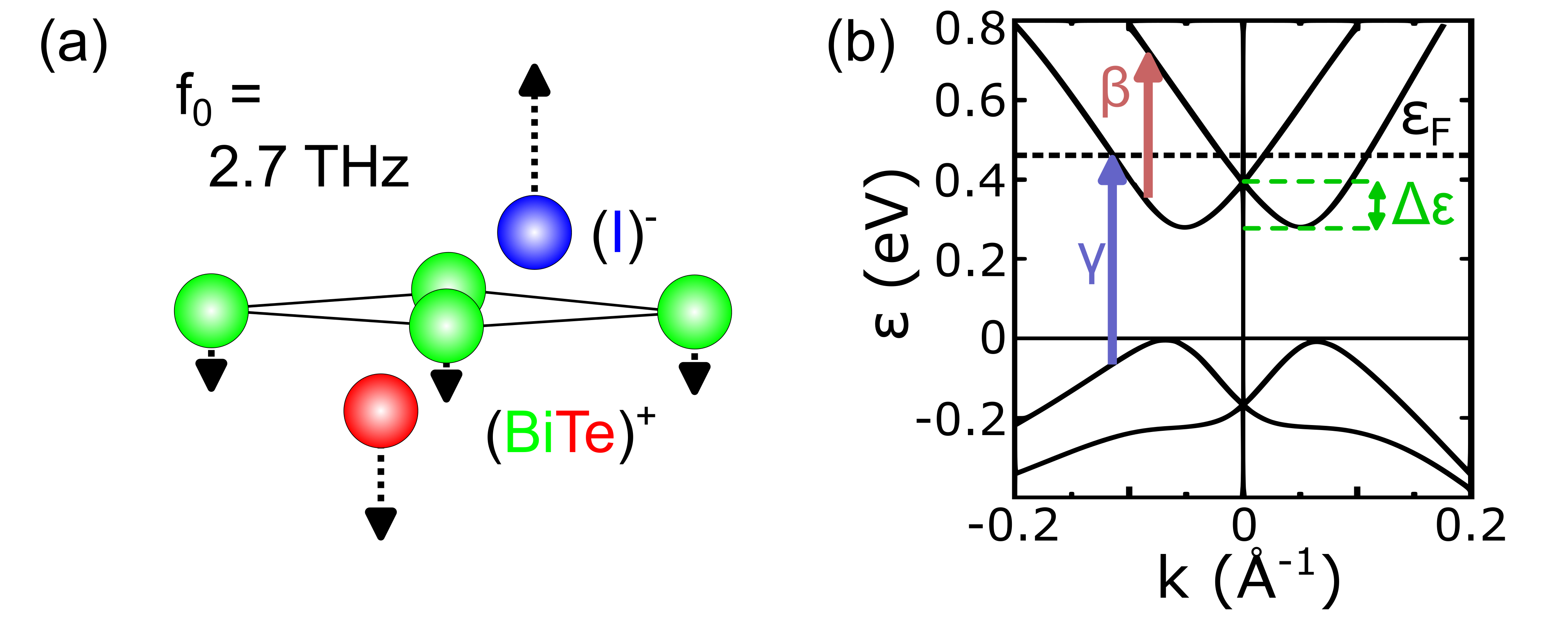}
\caption{\label{fig:BandNCrystal} (a) Atomic displacement and central frequency of an A\textsubscript{1} phonon mode in BiTeI \cite{Skl12}. (b) Rashba-split band structure of BiTeI (adapted from \cite{Bah11}, $\Delta \varepsilon=100$~meV).  The transition $\beta$ (red dashed line at 0.37~eV) represents the highest-energy intraband transition. The transition $\gamma$ (blue dashed line at 0.52~eV) is the lowest-energy interband transition corresponding to the electronic excitation from the valence band to a free state above the Fermi energy $\varepsilon_{\textnormal{F}}$ (dashed black line).}
\end{figure}

Our BiTeI specimen was grown by a modified Bridgman method using a rotating heat field \cite{Mak14}.
Its layered structure is built from covalently bound bilayers formed by (BiTe)\textsuperscript{+} which ionically bond to (I)\textsuperscript{-} and thus form trilayers.
These trilayers in turn are stacked by van der Waals forces.
We are investigating interband and intraband DECP in BiTeI via pump-probe time-domain reflection spectroscopy.
Therefore, we are interested in phonons with A\textsubscript{1} symmetry \cite{Zei92}.
As demonstrated by spontaneous Raman scattering \cite{Skl12}, BiTeI features an A\textsubscript{1} mode at a center frequency of 2.7~THz.
Its corresponding atomic displacement is shown in Fig. \ref{fig:BandNCrystal}~(a).
To perform our experiments, we employ two different amplified laser systems emitting fs pulses at kHz repetition rates (details in \cite{supp}).
A commercial system delivers laser pulses with photon energies tunable in the visible and near-infrared ranges (0.5~eV - 3~eV, 2~µm - 400~nm).
A self-built system \cite{Schoenfeld2024} generates pulses in the mid-infrared (0.18~eV, 7.0~µm).
Thus, we are able to employ photon energies both above and below the fundamental band gap of BiTeI (band-gap energy $\varepsilon_\textnormal{g} \approx 0.38$ eV \cite{Ish11}).
In all our experiments, the central wavelength of the probe beam is 1.2~µm (photon energy $\approx$ 1~eV). The reflected probe beam is recollected and detected by two photodiodes in a balanced detection scheme.
The signal is acquired via a digital lock-in method.
The sample was always at room temperature.
\\\\
Figure \ref{fig:BandNCrystal} (b) depicts the electronic band structure of BiTeI.
The compound is classified as a degenerate n-type semiconductor, implying that its Fermi energy $\varepsilon_\textnormal{F}$ lies above the conduction-band minimum \cite{Ish11} (black dashed line).
Because of the Rashba effect \cite{Ras15}, the dispersion of electrons with antiparallel spins is displaced in opposite directions from the center of the Brillouin zone.
The resulting energy splitting measured from the bottom of the conduction bands to their intersection is $\Delta \varepsilon\approx 100$~meV (green dashed lines) \cite{Ish11}.
Consequently, the absorption properties are characterized by two electronic transitions.
With $\beta$ we represent the highest-energy intraband transition between the Rashba-split conduction bands \cite{Dem12} (red arrow).
The lowest-energy interband transition, promoting electrons from the highest valence band to an unoccupied state above $\varepsilon_\textnormal{F}$ in the lowest conduction band, is denoted by $\gamma$ (blue arrow).
To determine the energies of these two transitions we measure the transmission spectrum of our sample (details in \cite{supp}) and obtain $\beta= 0.37$~eV (3.35~µm) and $\gamma=0.52$~eV (2.39~µm).
Depending on the crystal-growth conditions, the intrinsic charge-carrier density $n_\textnormal{i}$ of BiTeI can vary considerably. Since $\gamma$ depends on $\varepsilon_\textnormal{F}$, it is sensitive to $n_\textnormal{i}$.
Comparing the experimentally obtained value of $\gamma$ to the literature \cite{Lee11}, we estimate the intrinsic carrier density in our sample to be $n_\textnormal{i}=6\cdot10^{19}$~cm$^{-3}$.
\\\\
\section{\label{sec:Results}{Results and Discussion}}
\begin{figure*}[t]
\includegraphics{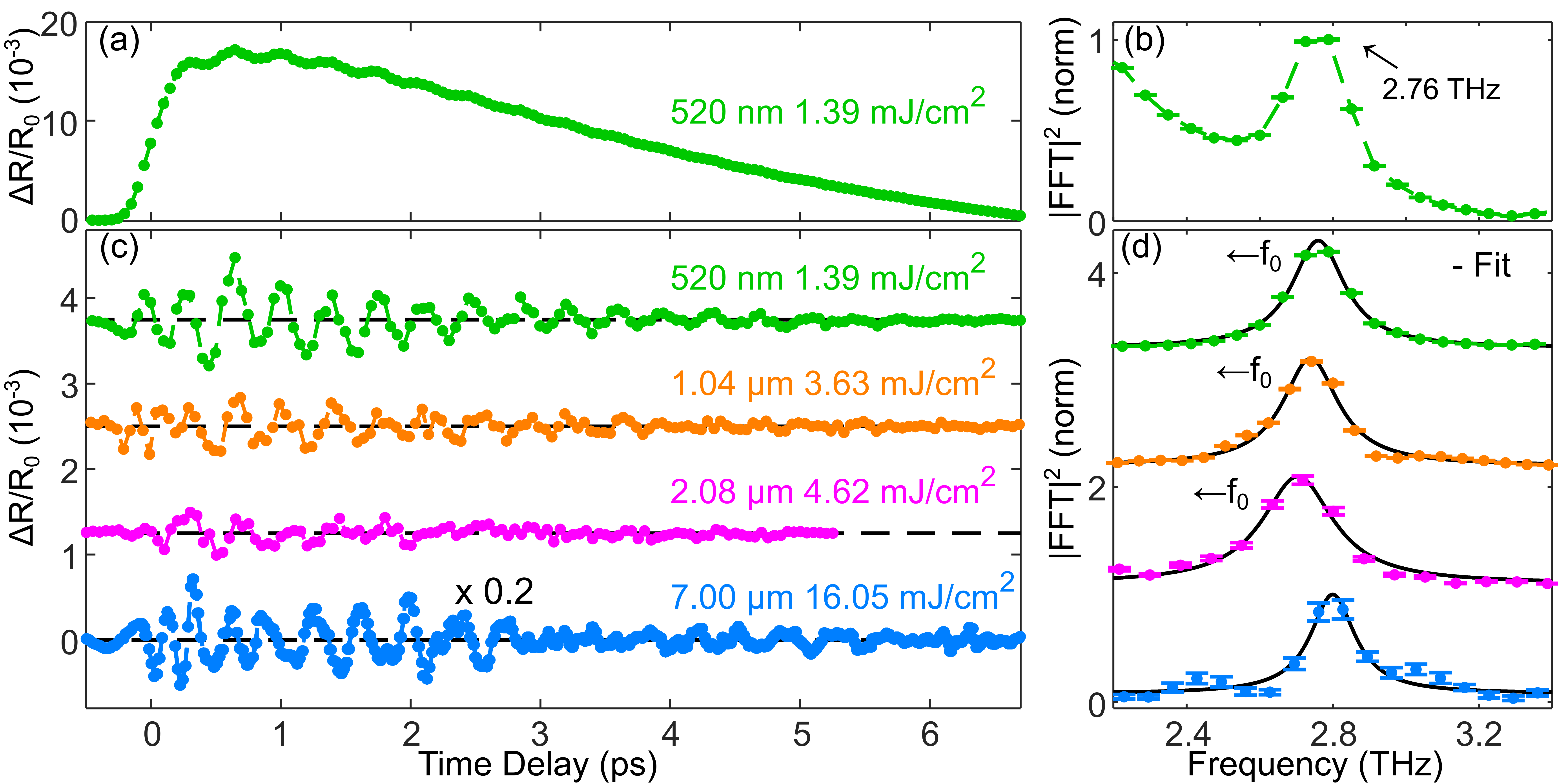}
\caption{\label{fig:ExampleMeas520nm}(a) Dynamics of the normalized reflectivity of BiTeI induced by 520-nm laser pulses. The fluence was set to 1.39~mJ/cm\textsuperscript{2}. (b) Power spectrum of the time trace shown in panel (a), obtained via Fourier transform. (c) Oscillatory component of the transient reflectivity for pump wavelengths spanning the visible, near-infrared and the mid-infrared spectral ranges after subtraction of the incoherent background. Note that for the 7.00-µm data (blue), the time trace was scaled with a factor of 0.2 for presentation purposes. (d) Power spectra of the time traces shown in panel (c). Each data set is normalized on its peak value at the respective central frequency $f_0$. The solid black lines represent a fit with a Lorentzian function (details in ~\cite{supp}). For interband excitation (520~nm - 2.08~µm) $f_0$ redshifts with increasing fluence (see explanation in the main text).}
\end{figure*}
\begin{figure}[t]
\includegraphics{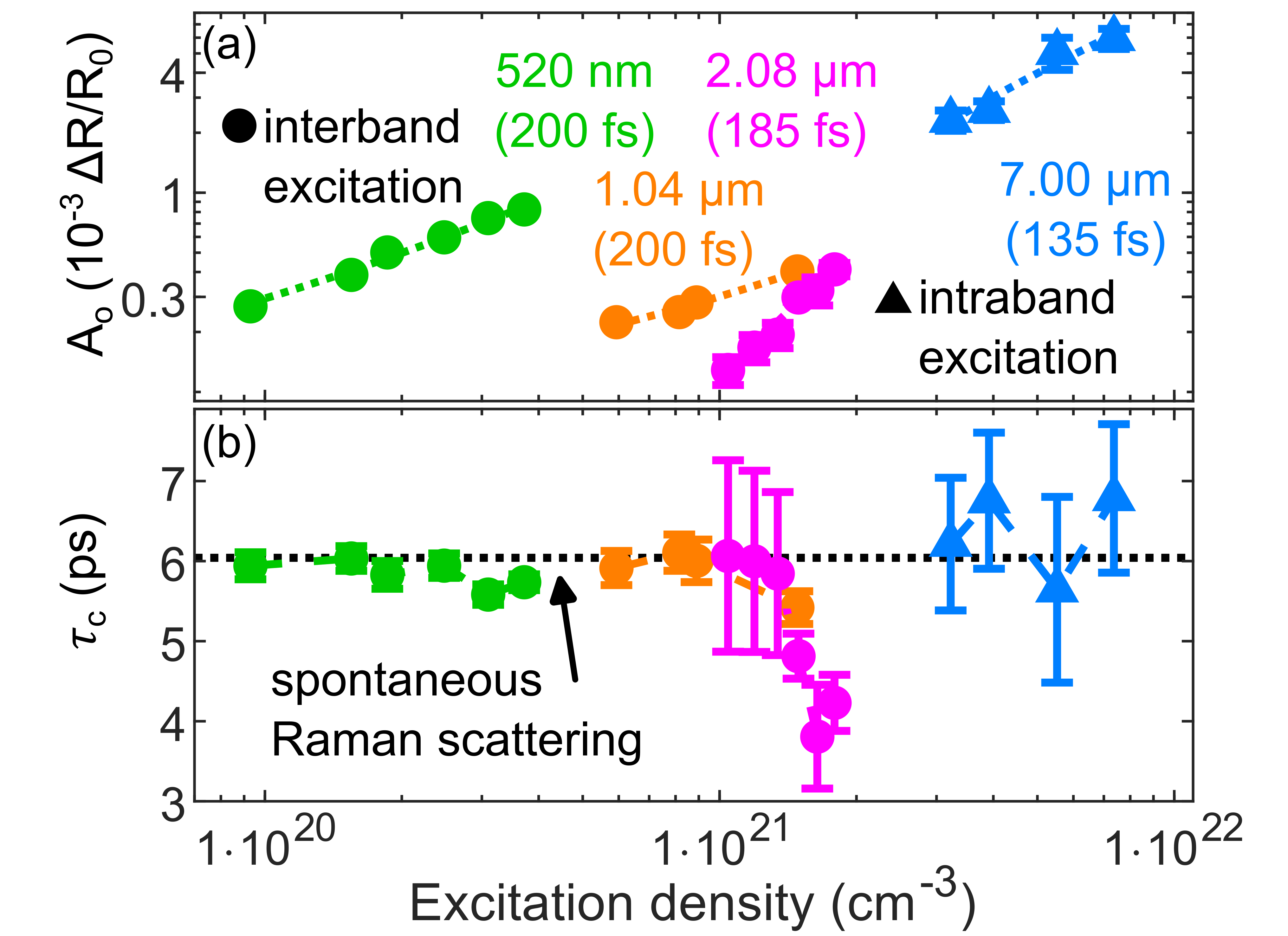}
\caption{\label{fig:flueDepParam}
Dependence of the phononic oscillation parameter on the excitation density.
Pump wavelengths (pulse durations) are indicated in the top panel. The error bars represent the 95$\%$ confidence interval.
(a) For each measurement series, the oscillation amplitude $A_\textnormal{o}$ shows a linear dependence on the excitation density. The data are plotted in a double-logarithmic scale for the sake of presentation, the dotted lines represent linear fits.
(b) Dependence of the coherence time $\tau_\textnormal{c}$ on the excitation density. The value obtained via Raman spectroscopy is shown by the dashed line \cite{Skl12}.}
\end{figure}

The main scientific goal of our work is to unravel the microscopic nature of DECP. 
We begin by characterizing the canonical regime of the displacive mechanism, in which the excitation of electronic interband transitions increases both $n_\textnormal{c}$ and $T_\textnormal{e}$.
\\\\
Figure \ref{fig:ExampleMeas520nm} (a) shows the dynamics of the normalized transient reflectivity $\Delta R/R_0$ induced by laser pulses with a central wavelength of 520~nm.
The pulse duration is 200~fs and the fluence is set to $\Phi=1.39$~mJ/cm\textsuperscript{2}.
The signal consists of two components, (i) harmonic oscillations with a period on the femtosecond timescale superimposed to (ii) an exponential decay on the picosecond timescale with an amplitude of $\Delta R/R_0\approx 1.5\cdot10^{-2}$.
Calculating the power spectrum via Fourier transform (Fig. \ref{fig:ExampleMeas520nm} (b)) reveals that the central frequency of the oscillations is 2.76~THz, a value consistent with the A\textsubscript{1} mode, whose atomic displacement corresponds to the out-of-phase oscillations of (BiTe)$^{+}$ and (I)$^{-}$  \cite{Skl12} (Fig \ref{fig:BandNCrystal} (a)).
We ascribe the exponential contribution to the signal to the thermalization of the photo-excited electrons with acoustic phonons \cite{Ket21}.
\\\\
Analyzing the isolated oscillatory component of $\Delta R/R_0$ (details in \cite{supp},  green circles in Fig. \ref{fig:ExampleMeas520nm} (c)) we obtain that the amplitude of the oscillations amounts to approximately $0.72\cdot 10^{-3}$.
To determine the coherence time $\tau_\textnormal{c}$, we calculate
the power spectrum shown by the green circles in Fig. \ref{fig:ExampleMeas520nm} (d) via Fourier
transform and fit it with a Lorentzian function (details in \cite{supp}).
We find $\tau_\textnormal{c}=(5.7\pm0.1)$~ps.
This result is in good agreement with the value obtained by spontaneous Raman scattering (6~ps)~\cite{Skl12}.
In contrast to the experiment reported in \cite{Skl12}, however, we set the fluence $\Phi$ to a value that significantly increases the free-carrier density $n_\textnormal{c}$.
Our analysis shows that this has no effect on $\tau_\textnormal{c}$, which is at first surprising since LO phonons usually interact strongly with free electrons \cite{Fro54}.
Driven by this finding, from $\Phi$ we calculate the excitation density, i.e. an estimate of the density of photons absorbed by electrons assuming that a single electron absorbs a single photon (details in \cite{supp}).
We highlight that in the interband regime the excitation density is equal to the increase of $n_\textnormal{c}$, i.e. the semiconductor is optically doped.
This procedure results in $n_\textnormal{c} \approx 4 \cdot 10^{20}$ cm$^{-3}$, a value almost one order of magnitude larger than the intrinsic charge-carrier density $n_\textnormal{i}$. In comparison to III-V semiconductors (GaAs, GaP, InAs, InSb) \cite{Kut92,Mis01,Yee02,Has99}, this behaviour is remarkable. For example, in GaAs free carriers with a density three orders of magnitude smaller strongly suppress the coherence time of the lattice oscillations \cite{Has99}. 
We attribute this difference to the van der Waals structure of BiTeI, due to which the electron wave vector is confined to the plane perpendicular to the stacking axis for states close to the conduction-band minimum.
The wave vector of LO phonons with A\textsubscript{1} symmetry, however, is parallel to the stacking.
Thus, the emission and absorption of such phonons by free electrons is suppressed due to the restrictions of energy and momentum conservation.
\\\\
Motivated by this observation, we quantitatively analyze the phononic robustness with respect to the excitation density.
For this purpose, we experimentally vary the pump fluence $\Phi$.
For each value of $\Phi$ we estimate the excitation density and analyze the amplitude $A_\textnormal{o}$ and the coherence time $\tau_\textnormal{c}$ of the oscillations from the Lorentz fit (details in \cite{supp}).
The results are shown in Fig. \ref{fig:flueDepParam} (a) and (b) by the green circles. The amplitude $A_\textnormal{o}$ of the phononic contribution to the signal (Fig. \ref{fig:flueDepParam} (a)) scales linearly with the excitation density, which is consistent with DECP \cite{MERL97}.
The value of $\tau_\textnormal{c}$ (Fig. \ref{fig:flueDepParam} (b)), on the other hand, shows no significant density-dependent variations up to $n_\textnormal{c}\approx4\cdot10^{20}$~cm$^{-3}$.
\\\\
Despite the scientific interest, we cannot further increase the fluence of the 520-nm pump beam, since the sample damage threshold would be surpassed.
We thus tune the central wavelength to 1.04~µm.
The difference between excitation at 520~nm and 1.04~µm is that the excited electrons have less excess energy in the latter case (i.e. the energy difference between the excited state and the bottom of the conduction band).
On the one hand, this allows us to further increase the fluence, as the damage threshold is higher compared to the 520-nm excitation.
Therefore, we can analyze $\tau_\textnormal{c}$ at even higher carrier densities.
However, this also means that the electronic subsystem reaches a lower value of $T_\textnormal{e}$ after internal thermalization.
This in turn allows us to investigate how the phononic amplitude $A_\textnormal{o}$ is affected when $T_\textnormal{e}$ is lower but $n_\textnormal{c}$ is comparable to the 520-nm excitation.
We will first address the latter aspect.
To begin with, with excitation at 1.04~µm we can increase the laser fluence to up to 3.63~mJ/cm\textsuperscript{2}, which corresponds to an excitation density of $1.5\cdot10^{21}$~cm\textsuperscript{-3}.
The oscillatory time trace of $\Delta R/R_0$ and the corresponding power spectrum measured at this highest fluence are shown in Fig. \ref{fig:ExampleMeas520nm} (c) and (d).
The results of the excitation-density dependence analysis (orange circles in
Fig. \ref{fig:flueDepParam} (a) and (b)) show that the amplitude $A_\textnormal{o}$ scales linearly with the excitation density as in the case of the 520-nm pump beam, however with smaller values.
In fact, considering comparable values of the excitation density ($4-6 \cdot 10^{20}$~cm\textsuperscript{-3}), $A_\textnormal{o}$ is 3.6 times bigger in the data set obtained with the 520-nm pump beam.
Thus, we infer that the rise in $T_\textnormal{e}$ is indeed a major factor for DECP.
With regards to $\tau_\textnormal{c}$ we observe a reduction of the parameter value by about 10\% at the highest excitation density ($1.5\cdot10^{21}$~cm\textsuperscript{-3}).
\\\\
To further investigate the matter we set the pump-pulse central wavelength to 2.08~µm (purple in Fig. \ref{fig:ExampleMeas520nm} (c)-(d) and Fig. \ref{fig:flueDepParam} (a) and (b)).
In comparison to the 1.04-µm excitation, the values of $A_\textnormal{o}$ in the 2.08-µm data are lower, as expected, but the slope is bigger.
This can be attributed to the different durations of the pump pulses, 200~fs at 1.04~µm compared to 185~fs at 2.08~µm.
The closer the temporal profile of the excitation pulses is to a delta function, the more efficient DECP is \cite{MERL97}.
\\\\
As far as $\tau_\textnormal{c}$ is concerned, a decrease down to 38$\%$ of the equilibrium value is observed at an excitation density of approximately $1.8 \cdot 10^{21}$ cm$^{-3}$ (see Fig. \ref{fig:flueDepParam} (b)), a value corresponding to $\Phi=4.62$~mJ/cm\textsuperscript{2}.
This decrease can also be seen in the time trace shown in Fig. \ref{fig:ExampleMeas520nm} (c) and in the broader power spectrum depicted in Fig. \ref{fig:ExampleMeas520nm} (d).
Conversely, we have demonstrated that BiTeI can maintain the coherence time of the A\textsubscript{1} phonon mode up to an optically doped carrier density of $n_\textnormal{c}\approx1\cdot10^{21}$~cm$^{-3}$.
\\\\
The eigenfrequency of A\textsubscript{1} mode redshifts as the excitation density increases (see Fig. \ref{fig:ExampleMeas520nm} (d)).
It decreases by approximately 2\% when comparing the values at the highest excitation densities between the 520-nm and 2.08-µm data.
This renormalisation of the lattice eigenfrequency stems from the screening of the ionic charges by the additional free charge carriers \cite{YU10}.
\\\\
So far, we have demonstrated that BiTeI is phononically robust and that the rise in $T_\textnormal{e}$ plays a major role in DECP.
However, pumping electronic interband transitions we could not completely disentangle the contribution of the rise in $T_\textnormal{e}$ from the increase of $n_\textnormal{c}$ to the phonon excitation.
Therefore, we employ pump pulses in the mid-infrared
range with a central wavelength of 7.00~µm.
The corresponding photon energy (0.18~eV) is not high enough to induce interband excitation and thus cannot change $n_\textnormal{c}$. 
However, mid-infrared pump pulses excite intraband transitions between the Rashba-split conduction bands (red arrow in Fig. \ref{fig:BandNCrystal} (b)).
Consequently, we can further increase $\Phi$ in comparison to previous measurements without damaging the sample.
In fact, we can employ fluences up to 16.1~mJ/cm\textsuperscript{2}, nearly four times higher than for 2.08-µm excitation and even one order of magnitude higher than for 520-nm excitation.
This translates to excitation densities up to approximately $7.4 \cdot 10^{21}$ cm$^{-3}$.
We emphasize that the interpretation of the excitation density in the intraband regime is restricted to density of absorbed photons and explicitly does not include the simultaneous increase of $n_\textnormal{c}$ as in the interband regime.
Nevertheless, the time-domain data (Fig. \ref{fig:ExampleMeas520nm} (c)) and the corresponding power spectrum (Fig. \ref{fig:ExampleMeas520nm} (d)) disclose that coherent phonons are excited even when $n_\textnormal{c}$ remains constant.
Combined with the conclusions of eletronic interband pumping, this compelling observation closes the fundamental issue standing for 30 years regarding the microscopy of DECP: the rise in $T_\textnormal{e}$ and not the increase of $n_\textnormal{c}$ is the dominant contribution to DECP.
\\\\
The amplitude retains its linear dependence on the fluence (blue triangulars in Fig. \ref{fig:flueDepParam} (a)).
From the absence of non-linearity in $A_\textnormal{o}$, we conclude that a single photon excites a single electron for intraband excitation as well.
A saturation of the amplitude due to bleaching of the electronic intraband transition could be expected, as the highest value for the excitation density is two orders of magnitude larger than $n_\textnormal{i}$.
However, photo-excited carriers in BiTeI scatter on a timescale shorter than 80~fs \cite{Mau11,Ket21}. As the duration of the 7.00-µm pump pulses is 135~fs, the conduction-band electrons are redistributed in momentum space within the duration of the optical excitation, thus preventing the absorption from saturating.
\\\\
In contrast to the interband regime, we observe that $\tau_\textnormal{c}$ shows no significant excitation-density-dependent variations (blue triangulars in Fig. \ref{fig:flueDepParam} (b)).
This can be attributed to the absence of additional scattering channels and energy dissipation mechanisms that typically arise with higher carrier densities.
Therefore, pumping electronic intraband transitions selectively in BiTeI allows to linearly increase $A_\textnormal{o}$ with the fluence while avoiding shortening $\tau_\textnormal{c}$.
Finally, we observe that in contrast to the interband regime the frequency of the A\textsubscript{1} phonon is not affected (see Fig. \ref{fig:ExampleMeas520nm} (d)).
As the mid-infrared pulses cannot increase $n_\textnormal{c}$, no screening of the ionic charges takes place.
\section{\label{sec:Conclusion}Conclusion}
We have demonstrated displacive excitation of coherent phonons in BiTeI.
By tuning the pump central wavelength in a broad spectral range spanning the visible, near-infrared and even mid-infrared regimes we observe DECP both for electronic interband and intraband transitions.
In the former scenario, we find that the phonon coherence time can be maintained for charge-carrier densities up to $n_\textnormal{c} = 1 \cdot 10^{21}$ cm$^{-3}$. 
This demonstrates a remarkable phononic robustness, since in III-V semiconductors the coherence time is already suppressed at values of $n_\textnormal{c}$ orders of magnitude lower.
The scenario of intraband transitions has been so far barely explored in terms of DECP.
Our data demonstrates that the photo-induced rise of $T_\textnormal{e}$ is the main microscopic contribution to DECP.
Furthermore, in this regime we can raise the phononic amplitude by increasing the fluence, while the coherence time is unaffected.
We observe that these results may be relevant for the development of optically driven ultrafast technology in quantum materials.
Schemes addressing the coherent structural manipulation of solids can be based on exciting electronic intraband transitions selectively, thus avoiding the additional scattering channels and energy dissipation induced by an increase of the carrier density.
Other routes, like Raman sum-frequency generation~\cite{Juraschek:2018ei} and nonlinear phononics~\cite{Foerst2011} are expected to induce the same phonon modes even in a nonlinear dynamical regime.
This is relevant with regard to the symmetry of the 2.7~THz mode, which may allow direct modification of the Rashba coupling \cite{Cre12,Che17,Kre22,Cio22,Mic22,Qu24}.
\\\\
\section*{Acknowledgements}
The exchange of samples took place in May 2021. The authors thank C. Beschle, and S. Eggert for their technical support. This work was supported by Deutsche Forschungsgemeinschaft (BO 5074/2-1), Davide Bossini acknowledges the support of the DFG program BO 5074/1-1.

\bibliography{apssamp}
\end{document}